\documentclass[12pt, letter, reqno]{amsart}
\addvspace{\medskipamount}
\makeatletter
\def\paragraph{\@startsection{paragraph}{4}%
  \z@{4mm}{-\fontdimen2\font}%
  {\normalfont\bfseries}}
\makeatother

\addvspace{\medskipamount}
\makeatletter
\def\subparagraph{\@startsection{subparagraph}{4}%
  \z@{4mm}{-\fontdimen2\font}%
  {\normalfont\bfseries}}
\makeatother

\usepackage{rotating}
\usepackage{array}
\usepackage{pdflscape}

\usepackage{amsfonts}       
\usepackage{amssymb}        
\usepackage{mathrsfs}		
\usepackage{amsmath}        
\usepackage{amsaddr}		
\usepackage{mathtools}		
\usepackage{graphicx} 		
\usepackage{wrapfig}		
\usepackage{alltt}          
\usepackage[pass]{geometry}	
\newgeometry{top=2.5cm,bottom=2.3cm,left=2.5cm,right=2.3cm}		
\usepackage{hyperref}		
\usepackage{url}			
\makeatletter
\def\url@leostyle{ \@ifundefined{selectfont}{\def\UrlFont{\sf}}{\def\UrlFont{\scriptsize\ttfamily}} } 
\makeatother  
\usepackage{endnotes}  		
\usepackage[perpage,symbol,multiple]{footmisc}  
\usepackage{epigraph}		

\usepackage{relsize}%

\makeatletter
\newcommand{\proofstep}[1]{%
  \par  
  \addvspace{\medskipamount}
  \textit{#1\@addpunct{.}}\enspace\ignorespaces
}
\makeatother

\newtheorem{example}{Example}
\newcommand{\etal}{\textit{et al.}}

\begin{document}
\title{A time-to-event three-outcome design for randomized phase II cancer trials}   
\author{Minghua Shan} 
\email{minghua.shan@bayer.com} 

\begin{abstract} 
Tumor response, a binary variable, has historically been the main measure of anti-tumor activity 
for many cancer phase II single-arm trials. Simon two-stage designs are often used.
Sargent \etal\  proposed a three-outcome trial design in this setting which requires
smaller sample sizes. For many new, molecularly targeted therapies, however, tumor response
may not be the most reliable endpoint for measuring anti-tumor activity.  Increasingly,
time-to-event endpoints, such as progression-free survival (PFS), are used in the phase II 
setting. When such endpoints are the primary measure of efficacy, a randomized concurrently
controlled study design is usually required.  Given limited resources for phase II, studies
are often underpowered with relatively large type I and II error rates, and it is sometimes
unavoidable to have a ``gray'' decision zone after phase II where a clear decision regarding
further development actions cannot be made without additional information.  Compared with an
underpowered standard two-outcome study, a three-outcome design prompts clinical trialists to
contemplate the likelihood of landing in the ``gray'' zone at the trial design stage and choose
study design parameters more appropriately.  We propose a three-outcome design, with or
without interim analyses, for randomized comparative phase II trials when a time-to-event
endpoint is used.

\medskip
\noindent \textbf{Key words:} three-outcome;
survival analysis; time-to-event data; sample size; group sequential; oncology phase II trial
\end{abstract}  

\maketitle
\pagestyle{plain}

\section{Introduction}
In oncology drug development, phase II studies are performed 
to screen for potentially efficacious treatments based on a surrogate endpoint which measures anti-tumor activity.
Single-arm study designs are commonly used when the primary outcome of interest is tumor response, a binary imaging-based variable. 
Simon two-stage\cite{simon} or Sargent \etal\  three-outcome\cite{sargent} procedures can be used
in this setting.
Three-outcome single-arm tumor response study design with continual monitoring is also available.\cite{shan} 

When an experimental drug is added to standard of care treatment and tumor response is considered
an appropriate endpoint for phase II evaluating the combination, randomized 
trials may be necessary when the standard of care treatment already produces a substantial 
tumor response rate.
Hong and Wang\cite{hong} extended the three-outcome single-arm design of Sargent \etal\  
to randomized comparative trials using tumor response as the primary measure of outcome. 

For many new classes of anti-cancer agents, such as molecularly targeted therapies, tumor response, which reflects tumor shrinkage, may not be the most reliable measure of anti-tumor activity. For many of these new treatments, their cytostatic properties may be a more complete measure of treatment effect.\cite{grayling}
For example, in a phase III trial, sorafenib produced only a 2\% tumor response rate in advanced hepatocellular carcinoma
but had a significant and meaningful overall survival benefit in the same patient population.\cite{llovet} Unlike tumor response, where tumors do not spontaneously shrink substantially without active treatment, time to tumor progression or death is part of disease's natural history.
Time-to-event data, such as 
progression-free survival (PFS), are usually not very informative without a concurrent control
because of the high level of variability and confounding with natural history of the disease.
When such an endpoint like PFS is used as the primary outcome measure,
randomization is recommended or even required regardless whether a single-agent or combination
regimen is tested.\cite{seymour}
These randomized phase II studies typically follow a standard time-to-event trial design but
often have higher error rates due to limited sample sizes in the phase II setting.
Compared with underpowered standard study designs, 
a three-outcome approach can be helpful in phase II because it prompts 
clinical trialists to consider the likelihood of inconclusive outcomes
at the trial design stage and facilitates proper choices of study design parameters. 

In this paper, we propose a three-outcome design for randomized cancer phase II trials
when a time-to-event endpoint is used as the primary measure of anti-tumor activity.
Group sequential analyses are possible when desired.
In Section~\ref{sec:tte3o}, we present the time-to-event three-outcome procedure. 
A group sequential three-outcome procedure is described in Section~\ref{sec:grpseq}. 
Some study design examples
are found in Section~\ref{sec:examples}. Section~\ref{sec:discussion} provides some discussions.

\section{Time-to-Event Three-Outcome Procedure}
\label{sec:tte3o}
Consider a two-arm randomized phase II cancer trial with a time-to-event variable, such as PFS, as the primary
efficacy endpoint. 
Let the two treatment groups be denoted by $i=0, 1$, where $0$ is the control and $1$ is the experimental arm.
Suppose patients are randomized $r:1$ to experimental or control treatment.
Let $j = 1,\dots,n$ represent the total of $n$ patients randomized to both study arms, and
let $D$ be the set of indices for all patients who are observed to be events.
We assume there are no tied event times.
Additionally, let $\lambda_i(t)$ be the hazard function for treatment group $i$, $i=0, 1$.
Define log hazard ratio $\theta=\theta(t)=log(\frac{\lambda_1(t)}{\lambda_0(t)})$. 
We assume $\theta$ is a constant (i.e., proportional hazards).
Hazard ratio (HR) is $exp(\theta)$.

Let the log-rank statistic,
\begin{equation}
\label{L_define}
L=\sum_{j \in D} (X_j - p_j ) \mathlarger{\mathlarger{/}}
\mathlarger{\mathlarger{[}}\sum_{j \in D} p_j(1-p_j)\mathlarger{\mathlarger{]}}^{1/2}, 
\end{equation}
where $X_j$ is an indicator for the experimental treatment group ($i=1$) for the $j^{th}$ patient,
and $p_j$ the proportion of patients at risk in treatment group $i=1$ just prior to the 
event from the $j^{th}$ patient.
Suppose $d$ is the total number of events observed in both study arms combined ($0<d \le n$).
When $d$ is large, $L$ is approximately normally distributed with variance 1 and mean\cite{schoenfeld}
\begin{equation}
\label{L_mean}
\frac{\theta \sqrt{rd}}{1+r}. 
\end{equation}
From \eqref{L_define} and \eqref{L_mean}, it immediately follows that
\begin{equation}
\label{theta_dist}
\hat{\theta} \equiv \frac{1+r}{\sqrt{rd}} L \sim N\left( \theta, \frac{(1+r)^2}{rd} \right).
\end{equation}

We would like to test the null hypothesis
\begin{equation*}
  H_0:\theta = \theta_0
\end{equation*}
versus the alternative hypothesis
\begin{equation*}
  H_1:\theta = \theta_1
\end{equation*}
where $0< \theta_1 < \theta_0$. Note that often $\theta_0 = 0$,
corresponding to hazard ratio of 1.0 or no treatment effect, and $\theta_1$ is a log
hazard ratio considered likely to predict a meaningful clinical benefit worth further
investigating post phase II.

\begin{figure}
  \includegraphics[width=1.0\linewidth,angle =0]{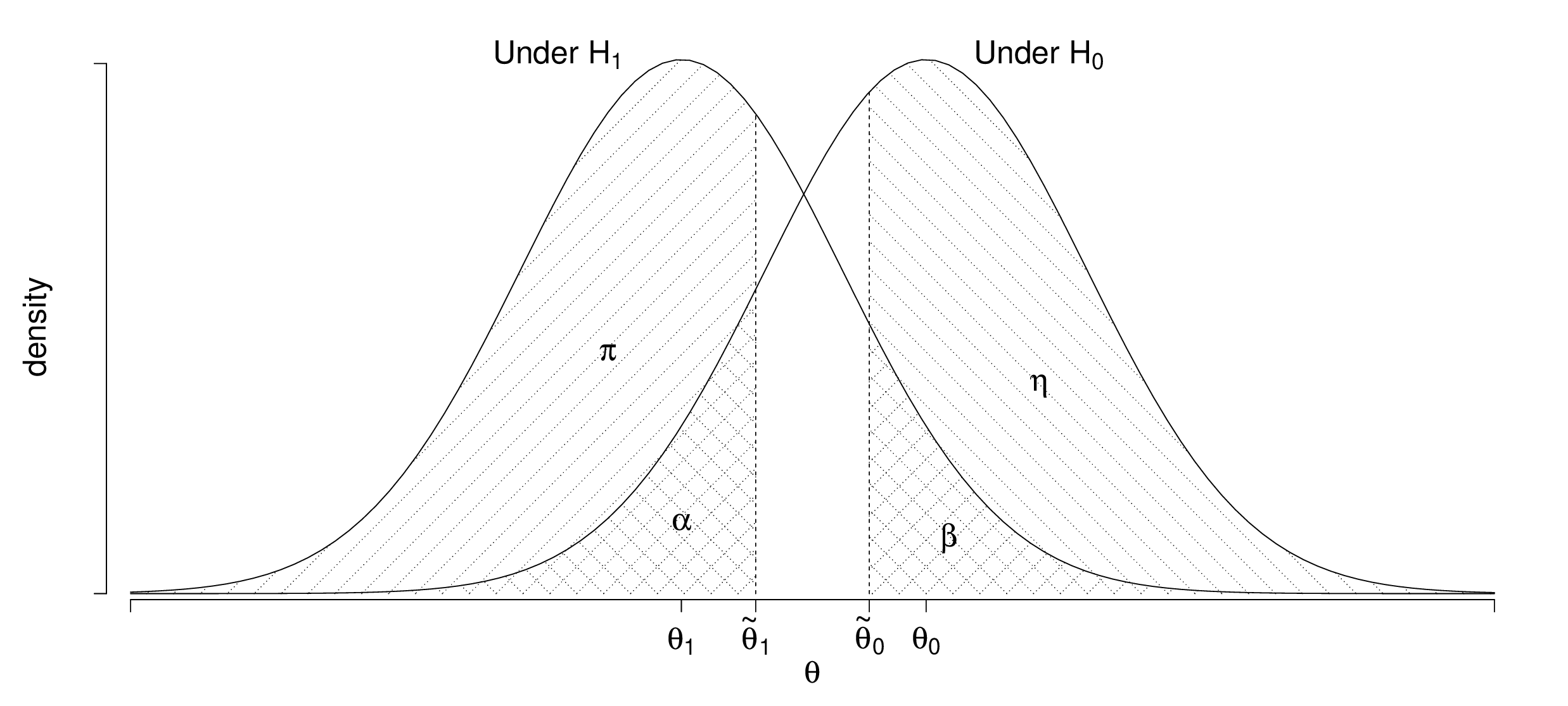}
  \caption{Probabilities for the time-to-event three-outcome design}
  \label{fig:plot}
\end{figure}

A three-outcome test procedure is defined by two decision boundaries, $\tilde{\theta}_0$ and $\tilde{\theta}_1$.
We reject $H_1$ if $\hat{\theta}>\tilde{\theta}_0$ and 
reject $H_0$ if $\hat{\theta}<\tilde{\theta}_1$.
We require that a test procedure meet the following conditions:
\begin{align}
\alpha & \ge P(\text{reject $H_0|H_0$ is true}),\label{alpha_cond}\\
\beta & \ge  P(\text{reject $H_1|H_1$ is true}),\label{beta_cond}\\
\eta & \le  P(\text{reject $H_1|H_0$ is true}),\label{eta_cond}\\
\pi & \le P(\text{reject $H_0|H_1$ is true}).\label{pi_cond}
\end{align}
That is, $\alpha$ and $\beta$ are the false positive and false negative error rates, $\pi$ is the probability of correctly rejecting $H_0$ (i.e., power), and $\eta$ is the probability of correctly rejecting $H_1$ (when $H_0$ is true).
Note that $1-\alpha-\eta$ and $1-\beta-\pi$ represent the probabilities of landing in the 
inconclusive or gray decision zone under $H_0$ and $H_1$, respectively.
Figure~\ref{fig:plot} shows the density function of $\hat{\theta}$ under $H_0$ and $H_1$, respectively, 
and areas under the curve corresponding to $\alpha, \beta, \eta$, and $\pi$.

From \eqref{theta_dist} and \eqref{alpha_cond}, setting \eqref{alpha_cond} to equality, we have
\begin{align*}
\alpha &= P(\hat{\theta}<\tilde{\theta}_1|\theta=\theta_0) \nonumber \\
       &= P\left(Z<\frac{(\tilde{\theta}_1 - \theta_0)\sqrt{rd}}{1+r}\right),
\end{align*}
where $Z \sim N(0,1)$. It follows that
\begin{equation}
\label{z_alpha}
z_\alpha = \frac{(\tilde{\theta}_1 - \theta_0)\sqrt{rd}}{1+r},
\end{equation}
where $z_\alpha$ is such that $P(Z<z_\alpha)=\alpha$.
Similarly, from \eqref{beta_cond}, \eqref{eta_cond}, and \eqref{pi_cond},
\begin{align}
-z_\beta &= \frac{(\tilde{\theta}_0 - \theta_1)\sqrt{rd}}{1+r}, \label{z_beta}\\
-z_\eta &= \frac{(\tilde{\theta}_0 - \theta_0)\sqrt{rd}}{1+r}, \label{z_eta}\\
z_\pi &= \frac{(\tilde{\theta}_1 - \theta_1)\sqrt{rd}}{1+r}, \label{z_pi}
\end{align}
with $z_\beta$, $z_\eta$, and $z_\pi$ similarly defined as $z_\alpha$.

Solving \eqref{z_beta} and \eqref{z_eta}, we obtain
\begin{equation}
\label{bnd0}
\tilde{\theta}_0 =\frac{z_\eta \theta_1 - z_\beta \theta_0}{z_\eta-z_\beta},
\end{equation}
and
\begin{equation*}
d_0 \equiv d =\frac{(z_\pi-z_\alpha)^2 (1+r)^2}{r(\theta_0-\theta_1)^2}.
\end{equation*}
Similarly, from \eqref{z_alpha} and \eqref{z_pi},
\begin{equation}
\label{bnd1}
\tilde{\theta}_1 =\frac{z_\pi \theta_0 - z_\alpha \theta_1}{z_\pi-z_\alpha},
\end{equation}
and
\begin{equation*}
d_1 \equiv d =\frac{(z_\eta-z_\beta)^2 (1+r)^2}{r(\theta_0-\theta_1)^2}.
\end{equation*}
If $d_0 > d_1$, we decrease false negative rate $\beta$ to make them equal. On the other hand, 
if $d_0 < d_1$, we increase power $\pi$ to make them the same value.
This gives us a unique test procedure with the minimum required number of events, $d$,
for a given set of specification. 
We believe lowering $\beta$ or increasing power $\pi$ is more appropriate in the phase II setting
as it reduces the likelihood of discarding a potentially efficacious drug without further investigation.
Note we may need to recalculate $\tilde{\theta}_0$ or $\tilde{\theta}_1$ based on \eqref{bnd0} or \eqref{bnd1}, respectively, using the updated $\beta$ or $\pi$ value.

Let $d \equiv d_0=d_1$, the minimum sample size required in terms of number of events.
The decision boundaries on the hazard ratio scale are $exp(\tilde{\theta}_0)$ and $exp(\tilde{\theta}_1)$
with the region in between being a gray or inconclusive zone.

\section{Group Sequential Three-Outcome Procedure}
\label{sec:grpseq}
Similar to standard (two-outcome) group sequential time-to-event study designs,
one or more interim analyses can be
incorporated into our three-outcome procedure based on the same $\alpha$- 
and $\beta$-spending approaches such as error spending functions.\cite{landemets}
With one interim analysis that has only a boundary for rejection of $H_1$,
the procedure is analogous to Sargent \etal's single-arm
two-stage three-outcome procedure for binary endpoint.\cite{sargent} 

Given the typically small sizes and relatively short durations of randomized,
especially three-outcome, phase II cancer trials, 
it is likely that most three-outcome time-to-event studies will have no or at most one interim analysis.
We illustrate our group sequential three-outcome procedure below using the case where
only one interim analysis is included. With only one interim analysis, obtaining sample size and 
boundaries numerically is not computationally intensive and direct numerical integration
and binary searches
can be used. Special considerations may be needed to achieve higher computational efficiency when the number of
analyses is large. See, for example, Chapter 19 of Jennison and Turnbull.\cite{jennison}

Suppose the number of events at the interim analysis is $d_1$ and $d_2$ additional events
are observed between the interim and final analyses. The total number of events at the final analysis
is $d=d_1+d_2$.
$t_1=d_1/d$ is the information fraction at which the interim analysis is to be performed.
Let $j = 1,\dots,n$ represent the total of $n$ patients randomized to both study arms.
Let $D_1$ be the set of indices for all patients who are observed to be events prior to the
interim analysis.  Similarly,
$D_2$ is the set of indices for all patients who are observed to be events between the interim
and final analyses.
Log-rank statistic $L$ is defined as in \eqref{L_define}.
Additionally, define
\begin{equation}
\label{L1_define}
L_1=\sum_{j \in D_1} (X_j - p_j ) \mathlarger{\mathlarger{/}}
\mathlarger{\mathlarger{[}}\sum_{j \in D_1} p_j(1-p_j)\mathlarger{\mathlarger{]}}^{1/2}, 
\end{equation}
and
\begin{equation}
\label{L2_define}
L_2=\sum_{j \in D_2} (X_j - p_j ) \mathlarger{\mathlarger{/}}
\mathlarger{\mathlarger{[}}\sum_{j \in D_2} p_j(1-p_j)\mathlarger{\mathlarger{]}}^{1/2}, 
\end{equation}
where $X_j$ and $p_j$ are as in Section~\ref{sec:tte3o}.
Similar to $\hat{\theta}$ in Section~\ref{sec:tte3o}, define
\begin{equation}
\label{theta_k}
\hat{\theta}_k = \frac{1+r}{\sqrt{rd_k}} L_k, \ k=1,2.
\end{equation}
Then $\hat{\theta}_k$ are independent and, for $d_k$ large, have approximately normal distributions
\begin{equation}
\label{theta_k_dist}
N\left( \theta, \frac{(1+r)^2}{rd_k}\right),\ k=1,2.
\end{equation}

When the ratio in size of risk set between the two study arms stays approximately constant,
$p_j \approx r/(1+r)$, for all $j \in D$, with $r:1$ randomization, and
\begin{equation*}
\mathlarger{\mathlarger{[}}\sum_{j \in D_k} p_j(1-p_j)\mathlarger{\mathlarger{]}}^{1/2}
\approx \frac{\sqrt{rd_k}}{1+r}, \ k=1,2.
\end{equation*}
We have
\begin{equation*}
L \approx \sqrt{\frac{d_1}{d}}L_1 + \sqrt{\frac{d_2}{d}}L_2 ,
\end{equation*}
and
\begin{equation*}
\hat{\theta} \approx \frac{d_1}{d}\hat{\theta}_1 + \frac{d_2}{d}\hat{\theta}_2 .
\end{equation*}

Let $\{\tilde{\theta}_0^{(1)}, \tilde{\theta}_1^{(1)}\}$ and
$\{\tilde{\theta}_0, \tilde{\theta}_1\}$ be boundaries for the interim and final analyses, respectively.
At the interim analysis, we reject $H_1$ if 
$\hat{\theta}_1 > \tilde{\theta}_0^{(1)}$ and reject $H_0$ if
$\hat{\theta}_1 < \tilde{\theta}_1^{(1)}$.
Further, let $\alpha_1$ and $\beta_1$ be the false positive and negative error rates for the 
interim analysis, which may be specified based on error spending functions or otherwise 
pre-specified based on practical considerations.
Our three-outcome group sequential procedure 
needs to meet these error spending values at
the interim analysis, performed at the specified information fraction $t_1$,
and satisfy study-wise error rates and 
power specifications, $\alpha, \beta, \pi$, and $\eta$ overall.

As $\hat{\theta}_1 \sim N\left( \theta, \frac{(1+r)^2}{rd_1}\right)$,
due to the requirements of $\alpha_1$ and $\beta_1$, we have
\begin{equation}
\label{bnd_i_1}
\tilde{\theta}_1^{(1)} = \theta_0 +\frac{z_{\alpha_1} (1+r)}{\sqrt{rd_1}}
\end{equation}
and
\begin{equation}
\label{bnd_i_0}
\tilde{\theta}_0^{(1)} = \theta_1 -\frac{z_{\beta_1} (1+r)}{\sqrt{rd_1}}.
\end{equation}

Given $\theta$, the probability of rejecting $H_0$ at the final analysis (without crossing
any boundary at the interim analysis) is
\begin{align}
A(\theta)\, = \,& P(\hat{\theta} < \tilde{\theta}_1,\,\tilde{\theta}_1^{(1)}<\hat{\theta}_1 < \tilde{\theta}_0^{(1)}
  |\theta) \nonumber\\
= \,& P(\frac{d_1}{d}\hat{\theta}_1 + \frac{d_2}{d}\hat{\theta}_2 < \tilde{\theta}_1,\,
  \tilde{\theta}_1^{(1)}<\hat{\theta}_1 < \tilde{\theta}_0^{(1)}|\theta)\nonumber\\
= \,& \int_{\tilde{\theta}_1^{(1)}}^{\tilde{\theta}_0^{(1)}}
\int_{-\infty}^{(d\tilde{\theta}_1 - d_1x_1)/d_2} f_1(x_1|\theta)f_2(x_2|\theta) \,dx_2 dx_1 
\nonumber \\
= \,& \int_{\tilde{\theta}_1^{(1)}}^{\tilde{\theta}_0^{(1)}} f_1(x_1|\theta)
F_2\left( \frac{d\tilde{\theta}_1 - d_1x_1}{d_2} \mathlarger{\mathlarger{|}} \theta\right) \,dx_1 
\label{alpha_f}
\end{align}
because of \eqref{theta_k_dist}, where $f_k(\cdot)$, $k=1,2$, are normal density functions for the 
distributions in \eqref{theta_k_dist} and $F_2(\cdot)$ is the cumulative probability function
corresponding to $f_2(\cdot)$.
Similarly,  the probability of rejecting $H_1$ at the final analysis is
\begin{align}
B(\theta)\,=\,&P(\hat{\theta} > \tilde{\theta}_0,\,\tilde{\theta}_1^{(1)}<\hat{\theta}_1 < 
 \tilde{\theta}_0^{(1)} |\theta) \nonumber\\
= \,& \int_{\tilde{\theta}_1^{(1)}}^{\tilde{\theta}_0^{(1)}} f_1(x_1|\theta)
 \left(1-F_2\left( \frac{d\tilde{\theta}_0 - d_1x_1}{d_2} \mathlarger{\mathlarger{|}} \theta\right)
 \right) \,dx_1. \label{beta_f}
\end{align}

To find the minimum sample size $d$ as well as interim and final analyses boundaries,
we perform a binary search on $d$ as follows. 
\begin{enumerate}
\item For any given value of $d$, $d_1 = t_1 d$. \label{step1}
\item Obtain interim analysis boundaries $\tilde{\theta}_0^{(1)}$ and $\tilde{\theta}_1^{(1)}$
 from \eqref{bnd_i_0} and \eqref{bnd_i_1}, respectively.
\item $A(\theta_0)$ is the final analysis specific false positive error rate.
  Using \eqref{alpha_f}, solve $A(\theta_0)=\alpha-\alpha_1$ for $\tilde{\theta}_1$ numerically.
\item $B(\theta_1)$ is the final analysis specific false negative error rate.
  Using \eqref{beta_f}, solve $B(\theta_1)=\beta-\beta_1$ for $\tilde{\theta}_0$ numerically.
\item Compute study-wise power $\pi(d)$ and correct negative rate $\eta(d)$
  for the current sample size $d$.
  \[\pi(d) = \int_{-\infty}^{\tilde{\theta}_1^{(1)}} f_1(x_1|\theta_1) dx_1 
  +A(\theta_1) \label{pi_d} \]
  \[\eta(d) = \int_{\tilde{\theta}_0^{(1)}}^{\infty} f_1(x_1|\theta_0) dx_1 
  +B(\theta_0) \label{eta_d} \]
\item If $\pi(d)<\pi$ or $\eta(d)<\eta$, increase $d$ and go to step~\ref{step1}. 
  If $\pi(d)>\pi$ and $\eta(d)>\eta$, decrease $d$ and go to step~\ref{step1}.
\end{enumerate}
We stop once reaching the desired precision.

\section{Examples}
\label{sec:examples}
\begin{samepage}
\begin{example}
\label{ex:1}
A three-outcome phase II study without interim analysis
\end{example}
\nopagebreak
As an example, we design a phase II cancer study comparing an experimental drug against a control treatment with PFS as the primary endpoint.
\end{samepage}
Randomization is 1:1 (i.e., $r=1$). We are interested
in testing the alternative hypothesis that the true hazard ratio is 0.65,
\begin{equation*}
H_1: \theta=\theta_1=log(0.65),
\end{equation*}
versus the null hypothesis of no activity,
\begin{equation*}
H_0: \theta=\theta_0=log(1)=0.
\end{equation*}
Additionally, we require false positive rate $\alpha$ and false negative rate $\beta$ to be $\le 0.15$, and power $\pi$ and correct negative rate $\eta$ $\ge 0.75$. This results in an approximately
$10\%$ inconclusive decision zone.

The number of events required, $d=63.1$, which we round up to $d=64$.
A sufficient number of patients needs
to be randomized so that $d=64$ events can be expected within a reasonable time frame.
Decision boundaries are $\tilde{\theta}_0 =-0.1686$ and $\tilde{\theta}_1 = -0.2591$ on the log-HR
scale, corresponding to hazard ratios of 0.8448 and 0.7717.

For comparison, a standard two-outcome procedure with $\alpha = \beta = 0.15$ requires $d=93$ events. Note that such a two-outcome procedure is equivalent to our three-outcome design
with $\pi = \eta = 0.85$ (that is, probabiltiy of inconclusive outcome is set to 0).

 Table~\ref{table1} displays additional study design scenarios showing events required and corresponding decision boundaries for the time-to-event three-outcome design.
See Appendix~\ref{appendix} for an R function for calculating minimum number of events needed and boundaries.
\begin{table}[h!]
\fontsize{9}{11}\selectfont
  \begin{center}
    \caption{Three-Outcome Examples}
    \label{table1}
    \begin{tabular}{l l l l l l r l l}
      \hline
      $HR_0$ &$HR_1$&$\alpha$ & $\beta$ & $\eta$& $\pi$&$d$&$exp(\tilde{\theta}_1)$
          &$exp(\tilde{\theta}_0)$\\
      \hline
1&0.5&0.1&0.1&0.8&0.8&38&0.6598&0.761\\
&&0.15&0.15&0.75&0.75&25&0.6606&0.7635\\
\hline
1&0.55&0.1&0.1&0.8&0.8&51&0.6984&0.79\\
&&0.15&0.15&0.75&0.75&33&0.6971&0.7907\\
&&0.15&0.15&0.7&0.7&28&0.6759&0.8202\\
\hline
1&0.6&0.1&0.1&0.8&0.8&70&0.7361&0.8178\\
&&0.15&0.15&0.75&0.75&45&0.7342&0.8178\\
&&0.15&0.15&0.7&0.7&38&0.7144&0.8435\\
&&0.2&0.2&0.7&0.7&29&0.7316&0.823\\
\hline
1&0.65&0.1&0.1&0.8&0.8&98&0.7719&0.8436\\
&&0.15&0.15&0.75&0.75&64&0.7717&0.8448\\
&&0.15&0.15&0.7&0.7&53&0.7522&0.8658\\
&&0.2&0.2&0.7&0.7&41&0.7688&0.8489\\
&&0.25&0.25&0.7&0.7&31&0.7848&0.8283\\
\hline
1&0.7&0.1&0.1&0.8&0.8&142&0.8065&0.8683\\
&&0.15&0.15&0.75&0.75&93&0.8066&0.8695\\
&&0.15&0.15&0.7&0.7&77&0.7896&0.8873\\
&&0.2&0.2&0.7&0.7&59&0.8032&0.8724\\
&&0.25&0.25&0.7&0.7&46&0.8196&0.8567\\
\hline
1&0.75&0.15&0.15&0.75&0.75&142&0.8403&0.893\\
&&0.15&0.15&0.7&0.7&118&0.8263&0.908\\
&&0.2&0.2&0.7&0.7&91&0.8382&0.8959\\
&&0.25&0.25&0.7&0.7&70&0.8511&0.8822\\
\hline
1&0.8&0.2&0.2&0.7&0.7&150&0.8716&0.9179\\
&&0.25&0.25&0.7&0.7&116&0.8823&0.9072\\
\hline
1.1&0.8&0.15&0.15&0.75&0.75&116&0.9074&0.9705\\
&&0.15&0.15&0.7&0.7&97&0.8912&0.9889\\
&&0.2&0.2&0.7&0.7&74&0.9045&0.9737\\
&&0.25&0.25&0.7&0.7&57&0.92&0.9573\\
\hline
1.2&0.8&0.1&0.1&0.8&0.8&110&0.9398&1.0221\\
&&0.15&0.15&0.75&0.75&72&0.9399&1.0236\\
&&0.15&0.15&0.7&0.7&60&0.9183&1.048\\
&&0.2&0.2&0.7&0.7&46&0.9363&1.0281\\
&&0.25&0.25&0.7&0.7&35&0.9553&1.0051\\
\hline
\multicolumn{8}{l}{\small Assumes 1:1 randomization.} & \\
    \end{tabular}
  \end{center}
\end{table}

\begin{samepage}
\begin{example}
\label{ex:2}
A three-outcome phase II study with one interim analysis
\end{example}
\nopagebreak
We design the same study as in Example~\ref{ex:1} but include one interim analysis
for potential rejection of $H_1$ only and thus stopping early for futility.
\end{samepage}
The interim analysis is planned at half of the total number of events for the final analysis
(i.e., $t_1=0.5$). Suppose that, of the study-wise $\beta$ of 0.15, $\beta$ level for the interim analysis is set at $\beta_1=0.05$.

The total number of events needed for the final analysis is $d=66$. Interim analysis is performed
at $d_1=33$. The interim analysis boundary for rejecting $H_1$ is $\tilde{\theta}_0^{(1)}=0.1419$,
corresponding HR of 1.1524. For the planned final analysis, 
$\tilde{\theta}_0^{(1)}=-0.1579$ and $\tilde{\theta}_1^{(1)}=-0.2513$, corresponding to hazard
ratios of 0.8539 and 0.7778, respectively.

\section{Discussion}
\label{sec:discussion}
Historically, especially with chemotherapies, single-arm cancer phase II trials have been commonly 
used when the primary efficacy endpoint is
tumor response, a binary variable. 
However, with the emergence of new classes of anti-cancer therapies, relying on tumor response as a 
main measure of anti-tumor activity has proven to be less reliable.

In recent times, randomized trials are becoming increasingly prevalent in oncology phase II drug 
development, particularly when the primary endpoint of interest is time to event.
Such endpoints are often based on imaging data, with progression-free survival being a 
commonly observed example.

Due to limited resources and in order to minimize the number of patients
exposed to potentially toxic but ineffective experimental treatments, study sizes in phase II are 
relatively small, leading to significant uncertainty in post-phase II development decisions.
Three-outcome study designs can facilitate consideration, at trial design stage,
of acceptable levels of uncertainty in the form of an inconclusive decision zone.
The proposed time-to-event three-outcome procedure for randomized trials can be used 
to design phase II studies when time-to-event endpoints such as PFS are the primary measure of activity.

Finally, our time-to-event three-outcome procedure resembles an interim analysis for
both efficacy and futility of a standard (two-outcome) group sequential time-to-event trial
testing a one-side hypothesis. At such an interim analysis, 
efficacy or futility may be declared depending on whether the corresponding
boundary is crossed; if neither boundary is crossed, the study continues to the next planned
analysis.
$\alpha$ and $\beta$ levels for the interim analysis are controlled
based on pre-specified error spending. However, there is no requirement on the
size of the ``continue'' or ``gray'' region for such an interim analysis.
Typically the probability of observing an interim
result in this region is high under both the null and alternative hypotheses.
Our three-outcome procedure requires pre-specification of the size of the ``gray'' region through
the other two parameters $\pi$ and $\eta$ because, unlike an interim analysis of a standard
group sequential design, we reach end of trial even if 
the result is inconclusive. This helps to control the probability of an inconclusive study
outcome to a level considered acceptable at the study design stage in the phase II setting.

\bibliographystyle{unsrt}
\bibliography{references}

\appendix
\section{R Code for Designing Randomized Time-to-Event Three-Outcome Trials}
\label{appendix}
{\fontsize{10}{11}\selectfont
\begin{alltt}
\input{appendix.tex}
\end{alltt}
}

\end{document}